\documentclass[reqno,12pt]{article}

\usepackage{amsmath}
\usepackage{latexsym}
\usepackage{amssymb}
\usepackage{hyperref}
\usepackage{graphicx}
\usepackage{epstopdf}
\usepackage{ifpdf}
\ifpdf
\fi

\newtheorem{theorem}{Theorem}[section]

\newtheorem{proposition}[theorem]{Proposition}
\newtheorem{lemma}[theorem]{Lemma}

\newtheorem{remark}[theorem]{Remark}
\newtheorem{assumption}[theorem]{Assumption}

\newcommand{\R}{{\mathbb R}}

\newcommand{\C}{{\mathbb C}}

\newcommand{\be}{\begin{equation}}
\newcommand{\ee}{\end{equation}}
\newcommand{\bea}{\begin{eqnarray}}
\newcommand{\eea}{\end{eqnarray}}
\newcommand{\ba}{\begin{array}}
\newcommand{\ea}{\end{array}}

\newcommand{\ol}{\overline}
\newcommand{\id}{\mathbb{I}}

\newcommand{\re}{\mathrm{Re}}

\newcommand{\eps}{\varepsilon}

\newcommand{\lam}{\lambda}
\newcommand{\Lam}{\Lambda}
\newcommand{\gam}{\gamma}
\newcommand{\Gam}{\Gamma}

\newcommand{\Om}{\Omega}
\newcommand{\dta}{\delta}

\newcommand{\tha}{\theta}

\numberwithin{equation}{section}

\begin{document}

\title{\bf The Ostrovsky-Vakhnenko equation on the half-line: a Riemann-Hilbert approach}
\author{Jian Xu\footnote{College of Science, University of Shanghai for Science and Technology, Shanghai 200093,
People's  Republic of China; email: jianxu@usst.edu.cn}~ and  Engui Fan\footnote{School of Mathematical Sciences, Key Laboratory of Mathematics for Nonlinear Science, Fudan University, Shanghai 200433,
People's  Republic of China; {\bf{correspondence author:}} email: faneg@fudan.edu.cn}\\
}


\maketitle

\begin{abstract}
We analyze an initial-boundary value problem for the Ostrovsky-Vakhnenko equation
 \[
 u_{xxt}-3u_{x}+3u_{x}u_{xx}+uu_{xxx}=0
 \]
 on the half-line. This equation can be viewed as the short wave model for the Degasperis-Procesi (DP) equation. We show
that the solution $u(x,t)$ can be recovered  from its initial and boundary values via the solution of a $3\times 3$
vector Riemann-Hilbert problem formulated in the complex plane of a spectral parameter $z$.

\end{abstract}

\section{\bf Introduction}

In this paper, we consider the partial differential equation (PDEs)
\be\label{sDPe}
u_{xxt}-3b u_{x}+3u_{x}u_{xx}+uu_{xxx}=0
\ee
where $b>0$ is a parameter and $u=u(x,t)$ is real-valued. This equation stems from the short-wave
limit of the Degasperis-Procesi equation (DP) \cite{dp}, which is a model of nonlinear shallow water waves:
\be\label{DPe}
u_t-u_{xxt}+3\omega u_x+4uu_{x}=3u_xu_{xx}+uu_{xxx}.
\ee
Indeed, introducing new space-time variables $(x',t')$ and a scaling of $u$ by
\[
x'=\frac{x}{\eps},\quad t'=\eps t,\quad u'=\frac{u}{\eps^2},
\]
where $\eps$ is a small positive parameter, then (\ref{sDPe}) is the leading term of (\ref{DPe}) as $\eps\rightarrow 0$. Thus,
equation (\ref{sDPe}) can be named as the ¡®short wave model of the Degasperis-Procesi
equation¡¯.

Equation (\ref{sDPe}) arises also in the theory of propagation of surface waves in
deep water, see \cite{klm}, as an asymptotic model for small-aspect-ratio waves.

For $\omega=0$, equation (\ref{sDPe}) reduces to the derivative Burgers equation
\[
(u_t+uu_x)_{xx}=0,
\]
whereas for $b=-\frac{1}{3}$, it reduces to the (differentiated) Vakhnenko equation \cite{v1,p}
\be\label{Ve}
(u_t+uu_x)_x+u=0.
\ee
Alternatively, (\ref{sDPe}) with $b=\frac{1}{3}$ reduces to (\ref{Ve}) after the change of variables $(u,t)\rightarrow (-u,-t)$.

Equation (\ref{Ve}) arises--and is known as the ¡®Vakhnenko equation¡¯¡ªin the context of
propagation of high-frequency waves in a relaxing medium \cite{v1,v2,v3}. On the other hand, being
written in the form
\be\label{Oe}
(u_t+c_0u_x+\alpha uu_x)_x=\gam u,
\ee
it is also called the ¡®reduced Ostrovsky equation¡¯ \cite{s}: it corresponds, in the case $\beta =0$, to
the equation
\be
(u_t+c_0u_x+\alpha uu_x+\beta u_{xxx})_x=\gam u,
\ee
that was derived by Ostrovsky in 1978 \cite{o}. Therefore, it is more correct to
name equation (\ref{Ve}) the ¡®Ostrovsky¨CVakhnenko equation¡¯ (OV), as it is proposed in \cite{bs}.

Well-posedness of the Cauchy problem for the Ostrovsky equation and its relatives
(reduced Ostrovsky equation, generalized Ostrovsky equation, etc) in Sobolev spaces has
been widely studied in the literature, using PDE techniques; see \cite{d,km,lm,ssk,vl}.

On the other hand, equation (\ref{sDPe}) is (at least, formally) integrable: it possesses a Lax pair
representation (see \cite{hw} appendix, (A1))
\begin{subequations}\label{laxpair}
\be\label{lax-x}
\psi_{xxx}=\lam(-u_{xx}+b)\psi
\ee
\be\label{lax-t}
\psi_t=\frac{1}{\lam}\psi_{xx}-u\psi_x+u_x\psi
\ee
\end{subequations}
where $\psi=\psi(x,t,\lam)$.

It is because of integrable, the Ostrovsky equation can be solved by the inverse scattering method. Recently, Boutet de Monvel and Shepelsky formulate a Riemann-Hilbert problem to the initial value problem, and they also consider the long-time asymptotic problem, see \cite{annebs}.

However, in many laboratory and field situations, the
wave motion is initiated by what corresponds to the imposition of
boundary conditions rather than initial conditions. This naturally
leads to the formulation of an initial-boundary value (IBV) problem
instead of a pure initial value problem.

In 1997, Fokas announced a new unified
approach for the analysis of initial-boundary value problems for
linear and nonlinear integrable PDEs \cite{f1,f2, f3}. The unified
method provides a generalization of the inverse scattering formalism
from initial value to initial-boundary value problems.  Over the
last almost two decades, it has been used to analyze boundary value
problems for several  important integrable equations with $2\times
2$ Lax pairs. Recently, Lenells develop a methodology for analyzing
initial-boundary value problems for integrable evolution equations
with Lax pairs involving $3\times 3$ matrices \cite{l3}. Then many other integrable evolution equations with $3\times 3$ Lax pairs are analyzed too, see \cite{l4,xf,xf2}.

In this paper,  we use Fokas and Lenells  method to analyze the initial-boundary value problem for the Ostrovsky-Vakhnenko equation (\ref{sDPe}) on the half-line domain, that is, in the domain
\be\label{xtdomain}
\Omega=\{(x,t)\in \R^2|0\le x<\infty, 0\le t<T\},
\ee
where $T<\infty$ is a given positive constant. Assuming that a solution exists, we show that $u(x,t)$ can be recovered from the initial and boundary values $u_0(x)$, $g_0(t),g_1(t),g_2(t)$ defined by
\be
\ba{llll}
u(x,0)=u_0(x),&0<x<\infty,&&\\
u(0,t)=g_0(t),&
 u_x(0,t)=g_1(t),&
  u_{xx}(0,t)=g_2(t),&0<t<T.
\ea
\ee
The main peculiarity compared with other applications of the approach of \cite{f1} is that the Lax pair involves $3\times 3$ matrices instead of $2\times 2$ matrices. This difference leads to some new challenges. Apart from the $3\times 3$ Lax pair, the spectral analysis of equation (\ref{sDPe}) on the half-line also presents some other peculiarities: (a) The presence of singularities in the Lax pair implies that it is necessary to introduce two sets of eigenfunctions. The eigenfunctions in the first set are well-behaved near $z=$. The eigenfunctions in the second set are well-behaved near $z=\infty$. Together these two sets of eigenfunctions can be used to formulate a Riemann-Hilbert problem. An analogous situation occurs in the analysis of DP equation on the half-line in \cite{l4}. (b) 
The basic matrix eigenfunctions which are natural candidates for the formulation of a Riemann-Hilbert problem, are difficult to recover the solution of the PDEs (\ref{OVe}). We overcome this problem by formulating an associated vector Riemann-Hilbert problem, for which it is much more easier to recover the solution of our problem in hand. (c) The formulation of the Riemann-Hilber problem depends, in addition to the variables $(x,t)$, on a function $y(x,t)$ which is unknown from the point of view
of the inverse problem. In order to obtain a Riemann-Hilbert problem whose jump matrix involves only known quantities, we have to
reparametrize the $x$ variable. This implies that we only obtain a parametric representation for the solution $u(x,t)$.

We will consider the initial-boundary value problems for (\ref{sDPe}) for which the initial and boundary values satisfy
\be\label{ibcon}
\ba{l}
-u_{0xx}(x)+b>0,\quad x\ge 0,\\
-g_2(t)+b>0,\quad 0\le t<T,
\ea
\ee
as well as
\be\label{dbcon}
g_0(t)\le 0,\quad 0\le t<T.
\ee
The assumptions in (\ref{ibcon}) imply that
\be
-u_{xx}(x,t)+b>0,\quad 0\le x<\infty, 0\le t<T.
\ee
The assumption (\ref{dbcon}) is used to ensure boundedness of certain eigenfunctions.

The organization of the paper is as follows.   In the following section 2,  we introduce two sets eigenfunctions.
In section 3, we derive expressions for the jump matrices in terms of suitable spectral functions. In section 4, we derive residue conditions for the pole singularities of the eigenfunctions. In section 5, we state our main result, see Theorem \ref{mainres}.

\section{\bf Spectral Analysis}

In this section, starting from the Lax pair of Ostrovsky-Vakhnenko equation, see (\ref{laxpair}), we  define analytic eigenfunctions
which are suitable for the formulation of a Riemann-Hilbert problem.

Without loss of generality, in what follows we assume that $b=1$. That is, we consider the the initial and boundary problems of the following equation
\be\label{OVe}
 u_{xxt}-3u_{x}+3u_{x}u_{xx}+uu_{xxx}=0
\ee

\subsection{Two sets of  eigenfunctions}
Let $z$ be the spectral parameter defined by $\lam=z^3$. The coefficients of the original Lax pair
(\ref{laxpair}) have singularities at $z=\infty$ and also at $z=0$. In order to have a good control on the
behavior of eigenfunctions at $z=\infty$ and at $z=0$ we introduce new forms of (\ref{laxpair}), the first
one appropriate at $z=\infty$, the second one at $z=0$.

Denote
\be
\Lambda(z)=\left(\ba{ccc}\lam_1(z)&0&0\\0&\lam_2(z)&0\\0&0&\lam_3(z)\ea\right),
\ee
where $\lam_j(z)=z\omega^j$, $j=1,2,3$, here $\omega=e^{\frac{2\pi i}{3}}$.

\subsubsection{Lax pair (well-controlled at $z=\infty$)}
Let $\Psi=\left(\ba{c}\psi\\\psi_x\\\psi_{xx}\ea\right)$, then the Lax pair (\ref{laxpair}) can be written as
\be\label{laxpair-matrix}
\left\{
\ba{l}
\Psi_x=\left(\ba{ccc}0&1&0\\0&0&1\\z^3q^3&0&0\ea\right)\Psi,\\
\Psi_t=\left(\ba{ccc}u_x&-u&z^{-3}\\1&0&-u\\-z^3uq^3&1-u_x\ea\right)\Psi,
\ea
\right.
\ee
where $q^3(x,t)=-u_{xx}(x,t)+1$.

Denote
\begin{subequations}\label{dpdef}
\be
D(x,t)=\left(\ba{ccc}q^{-1}(x,t)&0&0\\0&1&0\\0&0&q(x,t)\ea\right),
\ee
\be
P(z)=\left(\ba{ccc}1&1&1\\\lam_1(z)&\lam_2(z)&\lam_3(z)\\\lam^2_1(z)&\lam^2_2(z)&\lam^2_3(z)\ea\right)
\ee
\end{subequations}
\par
Setting $\tilde \Psi =P^{-1}D^{-1}\Psi$, the Lax pair (\ref{laxpair-matrix}) becomes
\be\label{laxpair-1}
\left\{
\ba{l}
\tilde \Psi_x-q\Lambda(z)\tilde \Psi=U\tilde \Psi,\\
\tilde \Psi_t+(uq\Lambda(z)-\Lam^{-1}(z))\tilde \Psi=V\tilde \Psi,
\ea
\right.
\ee
where
\begin{subequations}
\be
U=\frac{q_x}{3q}\left(\ba{ccc}0&1-\omega^2&1-\omega\\1-\omega&0&1-\omega^2\\1-\omega^2&1-\omega&0\ea\right),
\ee
\be
V=-uU+\frac{1}{3z}\left\{3\left(\frac{1}{q}-1\right)\id+\left(q^2-\frac{1}{q}\right)\left(\ba{ccc}1&1&1\\1&1&1\\1&1&1\ea\right)\right\}\left(\ba{ccc}\omega^2&0&0\\0&\omega&0\\0&0&1\ea\right)
\ee
\end{subequations}

Define $y(x,t)$ by
\be\label{ydef}
y(x,t)=\int_{(0,0)}^{(x,t)}q(x',t')(dx'-u(x',t')dt').
\ee
It is well-defined, because the conservation law
\be
q_t+(uq)_x=0,
\ee
implies that the integral in (\ref{ydef}) is independent of the path of integration.

Introducing $\tilde \Phi=\tilde \Psi e^{-y(x,t)\Lam(z)-t\Lam^{-1}(z)}$, we have
\be\label{laxpair-1n}
\left\{
\ba{l}
\tilde \Phi_x-[q\Lambda(z),\tilde \Phi]=U\tilde \Phi,\\
\tilde \Phi_t+[uq\Lambda(z)-\Lam^{-1}(z),\tilde \Phi]=V\tilde \Phi,
\ea
\right.
\ee
where brackets denote matrix commutator.

\begin{figure}[th]
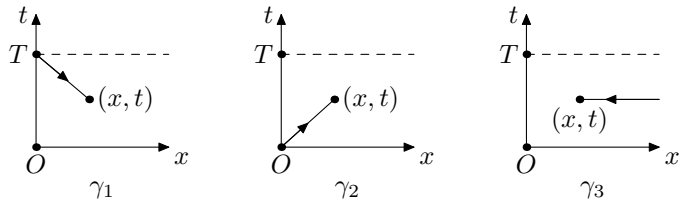

\centering
\includegraphics{OV--HL.1}\qquad
\includegraphics{OV--HL.2}\qquad
\includegraphics{OV--HL.3}
\caption{The three contours $\gam_1,\gam_2$ and $\gam_3$ in the $(x,t)-$domain.}\label{fig-1}
\end{figure}

We define three contours in $(x,t)-$domain, see Figure \ref{fig-1},

\begin{figure}[th]
\centering
\includegraphics{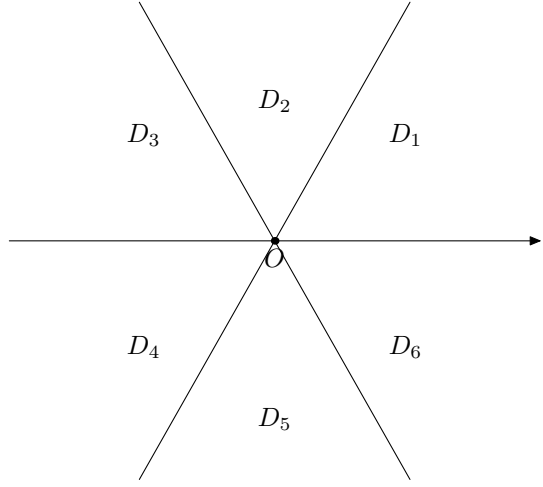}
\caption{The sets $\Omega_n$, $n=1,2,\dots,6$, which decompose the complex $z-$plane.}\label{fig-2}
\end{figure}

And we also denote six sets which decompose the complex $z-$plane, see Figure \ref{fig-2}
In these six sets $\Omega_n,n=1,2\dots,6$, the eigenvalues of $\Lam(z)$ and $\Lam^{-1}(z)$ has the following properties,
\be\label{eigendx}
\ba{ll}
D_1:&\{\re \lam_1<\re \lam_2<\re \lam_3,\quad \re \lam^{-1}_1<\re \lam^{-1}_2<\lam^{-1}_3\},\\
D_2:&\{\re \lam_1<\re \lam_3<\re \lam_2,\quad \re \lam^{-1}_1<\re \lam^{-1}_3<\lam^{-1}_2\},\\
D_3:&\{\re \lam_3<\re \lam_1<\re \lam_2,\quad \re \lam^{-1}_3<\re \lam^{-1}_1<\lam^{-1}_2\},\\
D_4:&\{\re \lam_3<\re \lam_2<\re \lam_1,\quad \re \lam^{-1}_3<\re \lam^{-1}_2<\lam^{-1}_1\},\\
D_5:&\{\re \lam_2<\re \lam_3<\re \lam_1,\quad \re \lam^{-1}_2<\re \lam^{-1}_3<\lam^{-1}_1\},\\
D_6:&\{\re \lam_2<\re \lam_1<\re \lam_3,\quad \re \lam^{-1}_2<\re \lam^{-1}_1<\lam^{-1}_3\},\\
\ea
\ee

The solutions of (\ref{laxpair-1n}) can be constructed as solutions of the
Fredholm integral equation
\be\label{zinftyMdef}
(\tilde \Phi_n(x,t,z))_{ij}=\dta_{ij}+\int_{\gam^{n}_{ij}}(e^{(y(x,t)-y(x',t'))\hat \Lam(z)+(t-t')\hat \Lam^{-1}(z)}(Udx'+Vdt')(x',t')\tilde \Phi(x',t',z))_{ij},\quad z\in \Omega_n,
\ee
where the contours $\gam_{ij}^n$, $n=1,2$, $i,j=1,2,3$ are defined by
\be\label{gamijndef}
\gam_{ij}^n=\left\{\ba{lclcl}\gam_1&if&\re \lam_i(z)<\re \lam_j(z)&and&\re \lam^{-1}_i(z)\ge\re \lam^{-1}_j(z),\\
\gam_2&if&\re \lam_i(z)<\re \lam_j(z)&and&\re \lam^{-1}_i(z)<\re \lam^{-1}_j(z),\\\gam_3&if&\re \lam_i(z)\ge\re \lam_j(z)&&.\\
\ea\right.\quad \mbox{for }\quad z\in \Omega_n.
\ee

\begin{remark}
For each $n=1,2,\dots,6$, the function $\tilde \Phi_{n}(x,t,z)$ is well-defined by equation (\ref{zinftyMdef}) for $z\in \bar \Omega_n$ and $(x,t)$ in the domain (\ref{xtdomain}). This is because of the definition of $\{\gam_j\}_1^3$ and (\ref{gamijndef}). The definition of $\{\gam_j\}_1^3$ implies that
\be\label{contourineql}
\ba{lll}
\gam_1:&{},&t-t'\le 0,\\
\gam_2:&y(x,t)-y(x',t')\ge 0,&t-t'\ge 0,\\
\gam_3:&y(x,t)-y(x',t')\le 0.&
\ea
\ee
The $(ij)$th entry of the integral equation (\ref{z0Mdef}) involves the exponential factor
\[
e^{(\lam_i(z)-\lam_j(z))(y(x,t)-y(x',t'))+(\lam^{-1}_i(z)-\lam^{-1}_j(z))(t-t')}.
\]
The definition of (\ref{gamijndef}) implies that this factor remains bounded for $z\in \Omega_n$ when integrated along the contour $\gam^n_{ij}$. In fact, it is similar to that of $\tilde \Phi_{0n}(x,t,z)$, except that since $y(x,t)-y(x',t')$ can take on both signs in the case of $\gam_1$, the exponential is not necessarily bounded for the integration along $\gam_1$. However, the matrices $(\gam^{n})_{ij}=\gam^n_{ij}$ for $n=1,2,\dots,6$ does not involve integration along $\gam_1$, so we can still conclude that $\tilde \Phi_n(x,t,z)$ is well-defined in each $\Omega_n$.
\end{remark}

\begin{proposition}
For any fixed point $(x,t)$, $\tilde \Phi_n(x,t,z)$ is bounded and analytic as
a function of $z\in \Omega_n, n=1,2,\dots,6$ away from a possible discrete set of singularities $\{z_j\}$ at which the
Fredholm determinant vanishes. Moreover, $\tilde \Phi_n(x,t,z)$ admits a bounded and continuous extension to $\bar \Omega_n$ and
\be\label{tphsnasy}
\tilde \Phi_n(x,t,z)=\id+O(\frac{1}{z}),\qquad z\rightarrow \infty,\quad z\in \Omega_n.
\ee
\end{proposition}

\subsubsection{Second Lax pair (well-controlled at $z=0$)}
Setting $\tilde \Psi_0=P^{-1}\Psi$, then the Lax pair (\ref{laxpair-matrix}) becomes
\be\label{laxpair-2}
\left\{
\ba{l}
\tilde \Psi_{0x}-\Lam(z)\tilde \Psi_0=U_0\tilde \Psi_0,\\
\tilde \Psi_{0t}-\Lam^{-1}(z)\tilde \Psi_0=V_0\tilde \Psi_0,
\ea
\right.
\ee
where
\begin{subequations}
\be
U_0=-\frac{zu_{xx}}{3}\left(\ba{ccc}\omega&0&0\\0&\omega^2&0\\0&0&1\ea\right)\left(\ba{ccc}1&1&1\\1&1&1\\1&1&1\ea\right)
\ee
\be
V_0=\frac{u_x}{3}\left(\ba{ccc}0&1-\omega^2&1-\omega\\1-\omega&0&1-\omega^2\\1-\omega^2&1-\omega&0\ea\right)-zu\left(\ba{ccc}\omega&0&0\\0&\omega^2&0\\0&0&1\ea\right)\left\{\id-\frac{u_xx}{3}\left(\ba{ccc}1&1&1\\1&1&1\\1&1&1\ea\right)\right\}
\ee
\end{subequations}

Setting $\tilde \Phi_0=\tilde \Psi_0 e^{-x\Lam(z)-t\Lam^{-1}(z)}$, then the Lax pair of $\tilde \Phi_0$ is
\be\label{laxpair-2n}
\left\{
\ba{l}
\tilde \Phi_{0x}-[\Lambda(z),\tilde \Phi_0]=U_0\tilde \Phi_0,\\
\tilde \Phi_{0t}-[\Lam^{-1}(z),\tilde \Phi_0]=V_0\tilde \Phi_0,
\ea
\right.
\ee
whose solutions can be constructed as solutions of the
Fredholm integral equation
\be\label{z0Mdef}
(\tilde \Phi_{0n}(x,t,z))_{ij}=\dta_{ij}+\int_{\gam^{n}_{ij}}(e^{(x-x')\hat \Lam(z)+(t-t')\hat \Lam^{-1}(z)}(U_0dx'+V_0dt')(x',t')\tilde \Phi_{0}(x',t',z))_{ij},\quad z\in \Omega_n,
\ee
where $\gam^{n}_{ij}$ are defined as (\ref{gamijndef}).

\begin{remark}
For each $n=1,2,\dots,6$, the function $\tilde \Phi_{0n}(x,t,z)$ is well-defined by equation (\ref{z0Mdef}) for $z\in \bar \Omega_n$ and $(x,t)$ in the domain (\ref{xtdomain}). This is because of the definition of $\{\gam_j\}_1^3$ and (\ref{gamijndef}). The definition of $\{\gam_j\}_1^3$ implies that
\be\label{contourineql}
\ba{ll}
\gam_1:&x-x'\ge 0,t-t'\le 0,\\
\gam_2:&x-x'\ge 0,t-t'\ge 0,\\
\gam_3:&x-x'\le 0.
\ea
\ee
The $(ij)$th entry of the integral equation (\ref{z0Mdef}) involves the exponential factor
\[
e^{(\lam_i(z)-\lam_j(z))(x-x')+(\lam^{-1}_i(z)-\lam^{-1}_j(z))(t-t')}.
\]
The definition of (\ref{gamijndef}) implies that this factor remains bounded for $z\in \Omega_n$ when integrated along the contour $\gam^n_{ij}$.
\end{remark}

\begin{proposition}
For any fixed point $(x,t)$, $\tilde \Phi_{0n}(x,t,z)$ is bounded and analytic as
a function of $z\in \Omega_n, n=1,2,\dots,6$ away from a possible discrete set of singularities $\{z_j\}$ at which the
Fredholm determinant vanishes. Moreover, $\tilde \Phi_{0n}(x,t,z)$ admits a bounded and continuous extension to $\bar \Omega_n$ and
\be\label{tphs0nasy}
\tilde \Phi_{0n}(x,t,z)=\id+z\tilde \Phi^{(1)}_0+z^2 \tilde \Phi^{(2)}_0+O(z^3),\qquad z\rightarrow 0,\quad z\in \Omega_n,
\ee
where
\be
\tilde \Phi^{(1)}_0=-\frac{1}{3}u_x \Gam,\quad \tilde \Phi^{(2)}_0=-\frac{1}{3}u\tilde \Gam,
\ee
with $\Gam=\left(\ba{ccc}\omega&\omega&\omega\\\omega^2&\omega^2&\omega^2\\1&1&1\ea\right)$ and $\tilde \Gam=[\tilde \Lam,\Gam]$, $\tilde \Lam=diag\{\omega,\omega^2,1\}$.
\end{proposition}

\subsubsection{Further properties of $\tilde \Phi_n$ and $\tilde \Phi_{0n}$}
Now, noticing that $\tilde \Phi_n$ and $\tilde \Phi_{0n}$ are related to the same
linear system of PDEs (\ref{laxpair-matrix}), tracing back the way that the differential equations for $\tilde \Phi_n$ and $\tilde \Phi_{0n}$
were derived, leads to the following proposition.

\begin{proposition}
The functions $\tilde \Phi_n$ and $\tilde \Phi_{0n}$ are related as follows,
\be\label{tworel}
\tilde \Phi_{n}(x,t,z)=P^{-1}(z)D^{-1}(x,t)P(z)\tilde \Phi_{0n}(x,t,z)e^{(x-y(x,t)\Lam(z))}.
\ee
where
$P(z)$ and $D(x,t)$ defined as (\ref{dpdef}), and $y(x,t)$ is defined as (\ref{ydef}).
\end{proposition}
Noticing that it is a straight calculus to show $P^{-1}(z)D^{-1}(x,t)P(z)$ is independent of $z$.

From (\ref{tphsnasy}), (\ref{tphs0nasy}) and (\ref{tworel}) we derive the following expansion of $\tilde \Phi_n(x,t,z)$ as $z\rightarrow 0$,
\be\label{tphsnasy0}
\ba{rcl}
\tilde \Phi_n(x,t,z)&=&P^{-1}(z)D^{-1}(x,t)P(z)\left(\id+z\left\{-\frac{u_x}{3}\Gam+(x-y)\tilde \Lam\right\}\right.\\
&&{}\left.+z^2\left\{-\frac{u}{3}\tilde \Gam-\frac{u_x}{3}(x-y)\Gam\tilde \Gam+\frac{(x-y)^2}{2}\tilde \Lam^2\right\}+O(z^3)\right)
\ea
\ee

\subsection{Symmetries}
\begin{proposition}\label{symcon}
$\tilde \Phi_n(x,t,z)$ satisfies the symmetry relations:
\begin{itemize}
  \item $\tilde \Phi_n(x,t,\omega z)=\mathcal{A}\tilde \Phi_n(x,t,z)\mathcal{A}^{-1}$, with $\mathcal{A}=\left(\ba{ccc}0&1&0\\0&0&1\\1&0&0\ea\right)$,
  \item $\tilde \Phi_n(x,t,z)=\mathcal{B}\ol{\tilde \Phi_n(x,t,\bar z)}\mathcal{B}^{-1}$, with $\mathcal{B}=\left(\ba{ccc}0&1&0\\1&0&0\\0&0&1\ea\right)=\mathcal{B}^{-1}$,
  \item $\tilde \Phi_n(x,t,z)=\mathcal{C}\ol{\tilde \Phi_n(x,t,\omega^2\bar z)}\mathcal{C}^{-1}$, with $\mathcal{C}=\left(\ba{ccc}0&0&1\\0&1&0\\1&0&0\ea\right)=\mathcal{C}^{-1}$.
  \item $\tilde \Phi_n(x,t,z)=\mathcal{D}\ol{\tilde \Phi_n(x,t,\omega\bar z)}\mathcal{D}^{-1}$, with $\mathcal{D}=\left(\ba{ccc}1&0&0\\0&0&1\\0&1&0\ea\right)=\mathcal{D}^{-1}$
\end{itemize}
The functions $\tilde \Phi_{0n}(x,t,z)$ satisfies the same symmetry relations.
\end{proposition}

\section{Sectionally Meromorphic Functions}

Let $\{\tilde \Phi_n\}_{1}^{6}$ and $\{\tilde \Phi_{0n}\}_1^6$ denote the eigenfunctions defined in Section 2. We have well control over $\tilde \Phi_n$ near $z=\infty$. On the other hand, we have well control over $\tilde \Phi_{0n}$ near $z=0$. Therefore, we will introduce a radius $R>0$ and formulate a Riemann-Hilbert problem by using the $\tilde \Phi_n$ for $|z|>R$ and the $\tilde \Phi_{0n}$ for $|z|<R$.

\begin{figure}[th]
\centering
\includegraphics{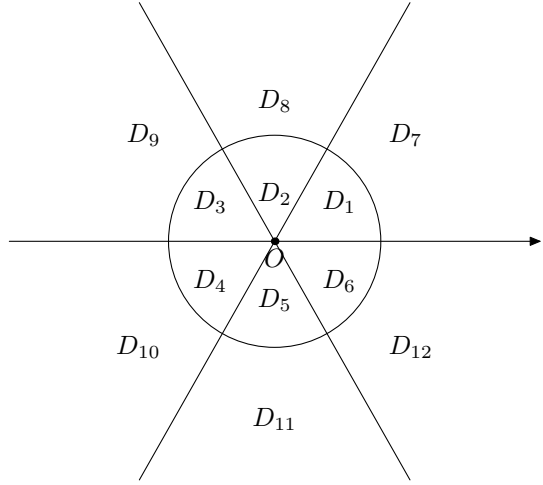}
\caption{The sets $\Omega_n$, $n=1,2,\dots,12$, which decompose the complex $z-$plane.}\label{fig-3}
\end{figure}

Define sets $\{D_n\}_1^{12}$ by (see Figure \ref{fig-3})

\be
\ba{ll}
D_n=\Omega_n \cap \{|z|>R\},&n=1,2,\dots,6,\\
D_{n+6}=\Omega_n \cap \{|z|<R\},&n=1,2,\dots,6.
\ea
\ee

Since the map $F: (x,t)\rightarrow (y,t), y=y(x,t)$ is a bijection from the domain (\ref{xtdomain}) onto $F(\Omega)$, we can define functions $\{M_n(y,t,z)\}_{1}^{12}$ for $(y,t)\in F(\Omega)$ by
\be\label{Mndef}
M_n(y,t,z)=\left\{
\ba{ll}
\tilde \Phi_{0n}(z,t,z)e^{(x-y)\Lam(z)},&z\in D_n,n=1,2,\dots,6,\\
P^{-1}(z)D(x,t)P(z)\tilde \Phi_{n}(x,t,z),&z\in D_n,n=7,8,\dots,12.
\ea
\right.
\ee
The $M_n$ defined in (\ref{Mndef}) are bounded and analytic on the complex $z-$plane. Reminding that the relation condition between the two eigenfunctions $\tilde \Phi_n$ and $\tilde \Phi_{0n}$, we need formulating the Riemann-Hilbert problem in terms of
the row vectors $\nu_n$ defined by
\be\label{nundef}
\nu_n(y,t,z)=\left(\ba{lll}1&1&1\ea\right)M_n(y,t,z),\quad z\in D_n,\quad n=1,2,\dots 12.
\ee

Let $M$ and $\nu$ denote the sectionally meromorphic functions on the complex $z-$plane which equal $M_n$ and $\nu_n$ respectively
for $z\in D_n$.

\begin{lemma}\label{symlemma}
The function $M$ obeys the  symmetries
\begin{itemize}
  \item $M(x,t,\omega z)=\mathcal{A}M(x,t,z)\mathcal{A}^{-1}$, with $\mathcal{A}=\left(\ba{ccc}0&1&0\\0&0&1\\1&0&0\ea\right)$,
  \item $M(x,t,z)=\mathcal{B}\ol{M(x,t,\bar z)}\mathcal{B}^{-1}$, with $\mathcal{B}=\left(\ba{ccc}0&1&0\\1&0&0\\0&0&1\ea\right)=\mathcal{B}^{-1}$,
  \item $M(x,t,z)=\mathcal{C}\ol{M(x,t,\omega^2\bar z)}\mathcal{C}^{-1}$, with $\mathcal{C}=\left(\ba{ccc}0&0&1\\0&1&0\\1&0&0\ea\right)=\mathcal{C}^{-1}$.
  \item $M(x,t,z)=\mathcal{D}\ol{M(x,t,\omega\bar z)}\mathcal{D}^{-1}$, with $\mathcal{D}=\left(\ba{ccc}1&0&0\\0&0&1\\0&1&0\ea\right)=\mathcal{D}^{-1}$
\end{itemize}
\end{lemma}
\textbf{Proof:}
This is a consequence of equation (\ref{Mndef}) and the symmetry properties of the $\tilde \Phi_n$ and the $\tilde \Phi_{0n}$. Noticing that
\be
P(z)=P(\omega z)\mathcal(A), \quad P(z)=\ol{P(\bar z)}\mathcal{B}.
\ee
And the other two symmetries can be obtained by these two symmetries.
$\Box$

\subsection{The jump matrices}

We define spectral functions $S_n(z)$ by
\be\label{Sndef}
S_n(z)=M_n(0,0,z),\qquad z\in D_n,\quad n=1,2,\dots,12.
\ee
The tracelessness of the matrices $\{U,V\}$ and $\{U_0,V_0\}$ implies that
\be
\det{S_n(z)}=1,\quad n=1,2,\dots,12.
\ee
The exponential factor $e^{(x-y)\Lam(z)}$ on the right-hand side of (\ref{Mndef}) has been included because it ensures that the jump matrices
introduced in the next proposition depend on $x$ only through the function $y(x,t)$.

\begin{proposition}
For each $n=1,2,\dots,12$, the function $\nu_n$ is bounded and analytic in $D_n$ (away from the possibly empty discrete
set $\{z_j\}$). Moreover, each $\nu_n$ has a continuous and bounded extension to $\bar D_n$. The function $\nu$ satisfies the jump conditions
\be\label{nujump}
\nu_n=\nu_mJ_{m,n},\quad z\in \bar D_n\cap \bar D_m,\quad n,m=1,\dots,12,\quad n\ne m,
\ee
where the jump matrix $J_{m,n}(y,t,z),J_{m,n}=J^{-1}_{n,m}$ is defined by
\be\label{Jdef}
J_{m,n}=e^{y\Lam(z)+t\Lam^{-1}(z)}(S^{-1}_mS_n)e^{-y\Lam(z)-t\Lam^{-1}(z)},\quad n,m\in \{1,\dots,12\}.
\ee
\end{proposition}
\textbf{Proof:}
The analyticity and boundedness properties of the $\nu_n$ follow from the properties of the $\tilde \Phi_n$ and the $\tilde \Phi_{0n}$ established
in Section 2.

From the definition of $\tilde \Phi_n$ (\ref{zinftyMdef}) and  $\tilde \Phi_{0n}$ (\ref{z0Mdef}), we can deduce that
\be\label{Mjump}
M_n=M_me^{y\Lam(z)+t\Lam^{-1}(z)}J(z)e^{-y\Lam(z)-t\Lam^{-1}(z)},
\ee
where $J(z)$ is a matrix independent of $(x,t)$.

Evaluation at $x=t=0$ yields $J=S^{-1}_mS_n$. Multiplying (\ref{Mjump}) by $\left(\ba{lll}1&1&1\ea\right)$ from the left, we obtain the jump condition (\ref{nujump})
with $J_{n,m}$ given by (\ref{Jdef}).
$\Box$

\section{Residue Conditions}

If the $\nu_n$ have pole singularities at some points $\{z_j\},z_j\in \C$, the Riemann-Hilbert problem needs to include the residue conditions at
these points. We will assume that all $z_j$ lie in the interiors of the sets $\{D_n\}_1^6$, singularities in the interiors of the sets $\{D_n\}_{7}^{12}$
can be avoided by choosing $R$ large enough. The residue conditions can be found
by relating the $M_n$ to another set of solutions of (\ref{laxpair-2}), denoted by $\{\mu_j\}_{1}^3$
, which are defined by
\be\label{mujdef}
\mu_j(x,t,z)=\id+\int_{\gam_j} e^{(x-x')\hat \Lam(z)+(t-t')\Lam^{-1}(z)}(Udx'+Vdt')(x',t')\mu_j(x',t',z),\quad j=1,2,3,
\ee
where $\{\gam_j\}_1^3$ are contours shown in Figure \ref{fig-1}.

\subsection{A matrix factorization problem}
Let us define the $3\times 3-$matrix value spectral functions $s(z)$ and $S(z)$ by
\begin{subequations}\label{sSdef}
\be\label{mu3mu2s}
\mu_3(x,t,z)=\mu_2(x,t,z)e^{(x \hat \Lam(z)+t\hat \Lam^{-1}(z))}s(z),
\ee
\be\label{mu1mu2S}
\mu_1(x,t,z)=\mu_2(x,t,z)e^{(x \hat \Lam(z)+t\hat \Lam^{-1}(z))}S(z).
\ee
\end{subequations}
Thus,
\be\label{sSmu3mu1}
s(z)=\mu_3(0,0,z),\qquad S(z)=\mu_1(0,0,z).
\ee

\begin{lemma}
Due to the symmetries of $M_n$, see Lemma \ref{symlemma}, we just need to calculate $S_1$. The $S_1$ defined in (\ref{Sndef}) can be expressed in terms of the entries of $s(z)$ and $S(z)$ as follows:
\be\label{Sn}
\ba{l}
S_1=\left(\ba{ccc}s_{11}&0&0\\
s_{21}&\frac{m_{33}(s)}{s_{11}}&0\\
s_{31}&\frac{m_{23}(s)}{s_{11}}&\frac{1}{m_{33}(s)}\ea\right),\\
\ea
\ee
where $m_{ij}$ denote that the $(i,j)-$th minor of $s$.
\end{lemma}
\textbf{Proof:}
Let $\gam_3^{X_0}$ denote the contour $(X_0,0)\rightarrow (x,t)$ in the $(x,t)-$plane, here $X_0>0$ is a constant.
We introduce $\mu_3(x,t,z;X_0)$ as the solution of (\ref{mujdef}) with $j=3$ and with the contour $\gam_3$ replaced
by $\gam_3^{X_0}$. Similarly, we define $M_1(x,t,z;X_0)$ as the solution of (\ref{Mndef}) with $\gam_3$ replaced
by $\gam_3^{X_0}$. We will first derive expression for $S_1(z;X_0)=M_1(0,0,z;X_0)$ in terms of $S(z)$ and
$s(z;X_0)=\mu_3(0,0,z;X_0)$. Then (\ref{Sn}) will follow by taking the limit $X_0\rightarrow \infty$.
\par
First, 
from the definition of $M_n$ (\ref{Mndef}), 
we have the following relations:
\be\label{MnRnSnTn}
\left\{
\ba{l}
M_1(y,t,z;X_0)=\mu_1(x,t,z)e^{(x\Lam(z)+t\Lam^{-1}(z))}R_1(z;X_0)e^{-y\Lam(z)-t\Lam^{-1}(z)},\\
M_1(y,t,z;X_0)=\mu_2(x,t,z)e^{(x\Lam(z)+t\Lam^{-1}(z))}S_1(z;X_0)e^{-y\Lam(z)-t\Lam^{-1}(z)},\\
M_1(y,t,z;X_0)=\mu_3(x,t,z;X_0)e^{(x\Lam(z)+t\Lam^{-1}(z))}T_1(z;X_0)e^{-y\Lam(z)-t\Lam^{-1}(z)}.
\ea
\right.
\ee
Then we get $R_1(z;X_0)$ and $T_1(z;X_0)$ are defined as follows:
\begin{subequations}\label{RnTnX0}
\be\label{RnX0}
R_1(z;X_0)=e^{-\hat \Lam^{-1}(z)T}M_1(y(0,T),T,z;X_0)e^{y(0,T)\Lam(z)},
\ee
\be\label{TnX0}
T_1(z;X_0)=e^{-\Lam(z)X_0}M_n(y(X_0,0),0,z;X_0)e^{y(X_0,0)\Lam(z)}.
\ee
\end{subequations}
The relations (\ref{MnRnSnTn}) imply that
\be\label{sSRnSnTn}
s(z;X_0)=S_1(z;X_0)T^{-1}_1(z;X_0),\qquad S(z)=S_1(z;X_0)R^{-1}_1(z;X_0).
\ee
These equations constitute a matrix factorization problem which, given $\{s(z),S(z)\}$ can be solved for the $\{R_1,S_1,T_1\}$.
Indeed, the integral equations (\ref{Mndef}) together with the definitions of $\{R_1,S_1,T_1\}$ imply that
\be
\left\{
\ba{lll}
(R_1(z;X_0))_{ij}=0&if&\gam_{ij}^1=\gam_1,\\
(S_1(z;X_0))_{ij}=0&if&\gam_{ij}^1=\gam_2,\\
(T_1(z;X_0))_{ij}=0&if&\gam_{ij}^1=\gam_3.
\ea
\right.
\ee
It follows that (\ref{sSRnSnTn}) are 18 scalar equations for 18 unknowns. By computing the explicit solution of this
algebraic system, we find that $\{S_1(z;X_0)\}$ are given by the equation obtained from (\ref{Sn}) by replacing
$\{S_1(z),s(z)\}$ with $\{S_1(z;X_0),s(z;X_0)\}$. Taking $X_0\rightarrow \infty$ in this equation,
we arrive at (\ref{Sn}).
$\Box$

\subsection{The residue conditions}
Since $\mu_2$ is an entire function, it follows that $M$ can only have singularities at the points
where the $S_1$ have singularities. We infer from the explicit formulas (\ref{Sn}) that the possible singularities
of $M$ are as follows:
\begin{itemize}
\item $[M]_2$ could have poles in $D_1$ at the zeros of $s_{11}(z)$;
\item $[M]_3$ could have poles in $D_1$ at the zeros of $m_{33}(z)$.
\end{itemize}
We denote the above possible zeros by $\{z_j\}_1^N$ and assume they satisfy the following assumption.
\begin{assumption}\label{assump}
We assume that
\begin{itemize}
\item $s_{11}(z)$ has $n_1$ possible simple zeros in $D_1$ denoted by $\{z_j\}_1^{n_1}$;
\item $m_{33}(z)$ has $N-n_1$ possible simple zeros in $D_1$ denoted by $\{z_j\}_{n_1+1}^{N}$;
\end{itemize}
and that none of these zeros coincide. Moreover, we assume that none of these functions have zeros on the boundaries
of the $D_n$'s.
\end{assumption}
We determine the residue conditions at these zeros in the following:
\begin{proposition}\label{propos}
Let $\{M_1\}$ be the eigenfunctions defined by (\ref{Mndef}) and assume that the set $\{z_j\}_1^N$ of singularities
are as the above assumption. Then the following residue conditions hold:
\begin{subequations}\label{resM}
\begin{align}
&\ba{l}{Res}_{z=z_j}[M]_2=\frac{m_{33}(s(z_j))}{\dot{ s_{11}(z_j)}s_{21}(z_j)}e^{\tha_{12}(z_j)}[M(z_j)]_1,\\
\quad 1< j\le n_1,z_j\in D_1\ea,\label{M12D1res}\\
&\ba{l}{Res}_{z=z_j}[M]_3=\frac{s_{11}(z_j)}{\dot{ m_{33}(s(z_j))}m_{23}(s(z_j))}e^{\tha_{23}(z_j)}[M(z_j)]_2,\\
\quad n_1< j\le N,z_j\in D_1\ea,\label{M13D1res}
\end{align}
\end{subequations}
where $\dot f=\frac{df}{dz}$, and $\tha_{ij}$ is defined by
\be\label{thaijdef}
\tha_{ij}(z)=(\lam_i-\lam_j)y+(\lam^{-1}_i-\lam^{-1}_j)t,\quad i,j=1,2,3.
\ee
\end{proposition}

\textbf{Proof:}
We will prove (\ref{M12D1res}).
From the relation
\be\label{M1S1}
M_1=\mu_2e^{(x\Lam(z)+t\Lam^{-1}(z))}S_1(z)e^{-y\Lam(z)-t\Lam^{-1}(z)},
\ee
For $i,j=1,2,3$, let $\tilde \tha_{ij}=(\lam_i-\lam_j)x+(\lam^{-1}_i-\lam^{-1}_j)t$. In view of the expressions for $S_1$ given in (\ref{Sn}), the three columns of (\ref{M1S1}) read:
\begin{subequations}
\begin{align}
&[M_1]_1e^{\lam_1(y-x)}=[\mu_2]_1s_{11}(z)+[\mu_2]_2e^{\tilde\tha_{21}}s_{21}(z)+[\mu_2]_3 e^{\tilde\tha_{31}}s_{31}(z),\label{M11}\\
&[M_1]_2e^{\lam_1(y-x)}=[\mu_2]_2\frac{m_{33}(s)}{s_{11}}+[\mu_2]_3e^{\tilde\tha_{32}}\frac{m_{23}(s)}{s_{11}},\label{M12}\\
&[M_1]_3e^{\lam_1(y-x)}=[\mu_2]_3\frac{1}{m_{33}(s)}.\label{M13}
\end{align}
\end{subequations}

In order to prove (\ref{M12D1res}), we suppose that $z_j\in D_1$ is a simple zero of $s_{11}(z)$. Solving (\ref{M11})
and (\ref{M13}) for $[\mu_2]_2$ and $[\mu_2]_3$ and substituting the result in to (\ref{M12}), we find
\[
[M_1]_2=\frac{m_{33}(s)}{s_{11}s_{21}}e^{\tha_{12}}[M_1]_1-\frac{m_{33}(s)}{s_{21}}e^{\tilde\tha_{12}+\lam_2(x-y)}[\mu_2]_1+\frac{m_{13}(s)m_{33}(s)}{s_{21}}e^{\tha_{32}}[M_1]_3.
\]
Taking the residue of this equation at $z_j$, we find the condition (\ref{M12D1res}) in the case
when $z_j\in D_1$. Similarly, we can get the equation (\ref{M13D1res}).
$\Box$

\section{\bf The Riemann-Hilbert problem}

The sectionally analytic function $\nu(y,t,z)$ defined in section 3 satisfies a Riemann-Hilbert problem
which can be formulated in terms of the initial and boundary values of $u(x,t)$. By solving this
Riemann-Hilbert problem, the solution of (\ref{OVe}) can be recovered in parametric form.
\begin{theorem}\label{mainres}
Suppose that $u(x,t)$ are a solution of (\ref{OVe}) on the half-line domain $\Om$ (\ref{xtdomain}) with sufficient
smoothness and decays as $x\rightarrow \infty$. Suppose that the initial and boundary values $\{u_0(x),g_0(t),g_1(t),g_2(t)\}$ defined in (\ref{inibouvalu}) satisfy the assumptions (\ref{ibcon}). Then $u(x,t)$ can be reconstructed from the initial
value and boundary values as follows.

Use the initial and boundary data to define $\{\tilde \Phi_n(0,0,z)\}_1^6$ and $\{\tilde \Phi_{0n}(0,0,z)\}_1^6$ via the integral equations (\ref{zinftyMdef}) and (\ref{z0Mdef}), respectively. Define spectral functions $S_n(z)$, $n=1,\dots,12$, by
\be
\ba{lll}
S_n(z)=\tilde \Phi_{0n-6}(0,0,z),&z\in D_n,&n=1,\dots,6,\\
S_n(z)=P^{-1}(z)D(0,0)P(z)\tilde \Phi_n(0,0,z),&z\in D_n,&n=7,\dots,12,
\ea
\ee
where $P(z)$ and $D(x,t)$ are defined in (\ref{dpdef}). Define
the jump matrices $J_{m,n}(y,t,z)$ in terms of the $S_n$ by equation (\ref{Jdef}). Define the spectral $s(z)$
and $S(z)$ by equation (\ref{sSdef}). Assume that the possible zeros $\{z_j\}_1^N$ of the functions $s_{11}(z),m_{33}(s(z))$ are as in assumption \ref{assump}.
\par
Then the solution $u(x,t)$ is given in parametric form by
\be
u(x,t)=\tilde u(y(x,t),t),
\ee
where
\begin{subequations}\label{solRHP}
\be\label{xsolRHP}
x(y,t)=y+\lim_{z\rightarrow 0}\frac{1}{z}\left(\nu^{(3)}(y,t,z)-1\right),
\ee
\be\label{usolRHP}
\tilde u(y,t)=\frac{\partial x(y,t)}{\partial t}.
\ee
\end{subequations}
with $\nu^{(3)}(y,t,z)$ is the third column of row-vector value function $\nu(y,t,z)$ which satisfies the following $3\times 3$ vector Riemann-Hilbert problem:
\begin{itemize}
\item $\nu(y,t,z)$ is sectionally meromorphic on the complex $z-$plane with jumps across the contour $\bar D_n\cap \bar D_m,n,m=1,2,\dots,12$, see Figure \ref{fig-2}.
\item Across the contour $\bar D_n\cap \bar D_m$, $\nu$ satisfies the jump condition (\ref{nujump})
\item $\nu(y,t,z)=\left(\ba{lll}1&1&1\ea\right)+O(\frac{1}{z}),\qquad z\rightarrow \infty$.
\item $\nu^{(2)}$ has simple poles at $z=z_j$, $1\le j\le n_1$. $\nu^{(3)}$ has simple poles at $z=z_j$, $n_1+1\le j\le N$. The associated residue condition  is showed in the following:
    \begin{subequations}\label{resnu}
\begin{align}
&\ba{l}{Res}_{z=z_j}\nu^{(2)}(y,t,z)=\frac{m_{33}(s(z_j))}{\dot{ s_{11}(z_j)}s_{21}(z_j)}e^{\tha_{12}(z_j)}\nu^{(1)}(y,t,z_j),\\
\quad 1< j\le n_1,z_j\in D_1\ea,\label{nu12D1res}\\
&\ba{l}{Res}_{z=z_j}\nu^{(3)}(y,t,z)=\frac{s_{11}(z_j)}{\dot{ m_{33}(s(z_j))}m_{23}(s(z_j))}e^{\tha_{23}(z_j)}\nu^{(1)}(y,t,z_j),\\
\quad n_1< j\le N,z_j\in D_1\ea,\label{nu13D1res}
\end{align}
\end{subequations}
where $\dot f=\frac{df}{dz}$, and $\tha_{ij}$ is defined by
\be\label{thaijdef}
\tha_{ij}(z)=(\lam_i-\lam_j)y+(\lam^{-1}_i-\lam^{-1}_j)t,\quad i,j=1,2,3.
\ee
\item For each zero $z_j\in D_1$, there are five additional points,
\[
\omega z_j,\quad \omega^2 z_j,\quad \bar k_j,\quad \omega \bar z_j,\quad \omega^2 \bar z_j,
\]
at which $\nu$ also has simple poles. The associated residues satisfy the residue conditions obtained from (\ref{resnu}) and the symmetries of Lemma \ref{symlemma}.
\end{itemize}
\end{theorem}
\textbf{Proof:}
The residue conditions (\ref{resnu}) are obtained by multiplying the conditions in (\ref{resM}) by $\left(\ba{lll}1&1&1\ea\right)$ from the left. It remains to prove (\ref{solRHP}). Equation (\ref{tphsnasy0}) implies that
\be
\nu(y,t,z)=\left(\ba{lll}\nu_1&\nu_2&\nu_3\ea\right)+O(z^3), \quad as \quad z\rightarrow 0,
\ee
where
\begin{subequations}
\be
\nu_1=\left(1+z\omega (x-y)+\frac{1}{2}z^2\omega^2(x-y)^2\right),
\ee
\be
\nu_2=\left(1+z\omega^2 (x-y)+\frac{1}{2}z^2\omega(x-y)^2\right),
\ee
\be\label{nu3}
\nu_3=\left(1+z (x-y)+\frac{1}{2}z^2(x-y)^2\right),
\ee
\end{subequations}

Thus, from the equation (\ref{nu3}) 
we can get (\ref{solRHP}).
$\Box$

{\bf Acknowledgements}
This work of Xu was supported by Shanghai Sailing Program
supported by Science and Technology Commission of Shanghai Municipality
under Grant NO.15YF1408100, Shanghai youth teacher assistance program NO.ZZslg15056 and the Hujiang Foundation of China (B14005). Xu also wants thank Shanghai Center for Mathemathical Sciences(SCMS) for honest inviting, most of the work is started in there. Fan was support by grants from the National
Science Foundation of China (Project No.10971031; 11271079; 11075055).


\begin{thebibliography}{XXXX}

\bibitem{dp} Degasperis A and Procesi M 1999 Asymptotic Integrability, Symmetry and Perturbation Theory
(Rome, 1998) (River Edge, NJ: World Science) pp 23-37.

\bibitem{klm} Kraenkel R A, Leblond H and Manna M A 2014 An integrable evolution equation for surface
waves in deep water, {\em J. Phys. A: Math. Theor.} 47 025208.

\bibitem{v1} Vakhnenko V O 1992 Solitons in a nonlinear model medium, {\em J. Phys. A: Math. Gen.} 25 4181-7.

\bibitem{p} Parkes E J 1993 The stability of solutions of Vakhnenkos equation, {\em J. Phys. A: Math. Gen.} 26
6469-75.

\bibitem{v2} Vakhnenko V O 1997 The existence of loop-like solutions of a model evolution equation, {\em Ukr. J.
Phys.} 42 104-10.

\bibitem{v3} Vakhnenko V O 1999 High-frequency soliton-like waves in a relaxing medium {\em J. Math. Phys.} 40
2011-20

\bibitem{s} Stepanyants Y A 2006 On stationary solutions of the reduced Ostrovsky equation: periodic waves,
compactons and compound solitons, {\em Chaos Solitons Fractals} 28 193-204.

\bibitem{o} Ostrovsky L A 1978 Nonlinear internal waves in a rotating ocean, {\em Oceanology} 18 181-91.

\bibitem{bs} Brunelli J C and Sakovich S 2013 Hamiltonian structures for the Ostrovsky¨CVakhnenko equation, {\em
Commun. Nonlinear Sci. Numer. Simul.} 18 56-62.

\bibitem{d} Davidson M 2013 Continuity properties of the solution map for the generalized reduced Ostrovsky
equation, {\em J. Differ. Equ.} 252 3797-815.

\bibitem{km} Khusnutdinova K R and Moore K R 2011 Initial-value problem for coupled Boussinesq equations
and a hierarchy of Ostrovsky equations, {\em Wave Motion} 48 738-52.

\bibitem{lm} Linares F and Milan¨¦s A 2006 Local and global well-posedness for the Ostrovsky equation,
{\em J. Differ. Equ.} 222 325-40.

\bibitem{ssk} Stefanov A, Shen Y and Kevrekidis P G 2010 Well-posedness and small data scattering for the
generalized Ostrovsky equation, {\em J. Differ. Equ.} 249 2600-17.

\bibitem{vl} Varlamov V and Liu Y 2004 Cauchy problem for the Ostrovsky equation, {\em Discrete Contin. Dyn.
Syst.} 10 731-53.

\bibitem{hw} Hone A N W and Wang J P 2003 Prolongation algebras and Hamiltonian operators for peakon
equations, {\em Inverse Problems} 19 129-45.

\bibitem{annebs} Boutet de Monvel A and Shepelsky D 2015 The Ostrovsky-Vakhnenko equation by a Riemann-Hilbert approach, {\em J. Phys. A: Math. Theor.} 48 035204.

\bibitem{f1} Fokas AS 1997 A unified transform method for solving linear and certain nonlinear PDEs,
{\em Proc. R. Soc. Lond. A} 453 1411-1443.

\bibitem{f2} Fokas AS 2002 Integrable nonlinear evolution equations on the half-line,
{\em Commun. Math. Phys.} 230 1-39.

\bibitem{f3} Fokas AS 2008 A unified approach to boundary value problems, in: CBMS-NSF Regional Conference
Series in Applied Mathematics, SIAM.

\bibitem{l3} Lenells J 2012 Initial-boundary value problems for integrable evolution equations with $3\times 3$
Lax pairs, {\em Phys D } 241 857-875.

\bibitem{l4} Lenells J 2013 The Degasperis-Procesi equation on the half-line, {\em Nonlinear Analysis }
76 122-139.

\bibitem{xf} Xu J and Fan E 2013 The unified transform method for the Sasa-Satsuma equation on the half-line, {\em Proc. R. Soc. A,} 469.

\bibitem{xf2} Xu J and Fan E 2014 The three wave equation on the half-line, {\em Physics Letters A,} 378 26-33.

\bibitem{gp} Grimshaw R and Pelinovsky D 2014 Global existence of small-norm solutions in the reduced
Ostrovsky equation, {\em Discrete Contin. Dyn. Syst.} 34 557-66.


\end{thebibliography}
\end{document}